\documentclass{ws-ijmpe}
\usepackage{amssymb}

\begin{document}

\markboth{Renxin Xu}{Pulsars: Gigantic Nuclei}

\catchline{}{}{}{}{}

\title{Pulsars: Gigantic Nuclei}

\author{Renxin Xu}

\address{School of Physics and State Key Laboratory of Nuclear Physics and
Technology,\\
Peking University, Beijing 100871, China\\
{\tt r.x.xu@pku.edu.cn}}


\maketitle


\begin{abstract}
What is the real nature of pulsars? This is essentially a question
of the fundamental strong interaction between quarks at low-energy
scale and hence of the non-perturbative quantum chromo-dynamics, the
solution of which would certainly be meaningful for us to understand
one of the seven millennium prize problems (i.e., ``Yang-Mills
Theory'') named by the Clay Mathematical Institute.
After a historical note, it is argued here that a pulsar is very
similar to an extremely big nucleus, but is a little bit different
from the {\em gigantic nucleus} speculated 80 years ago by L.
Landau.
The paper demonstrates the similarity between pulsars and gigantic
nuclei from both points of view: the different manifestations of
compact stars and the general behavior of the strong interaction.
\end{abstract}

\vspace{0.2cm}
\noindent%
{\bf 1. What is a Gigantic Nucleus?}

Let's begin with the {\it little} nucleus in normal matter.
Atoms were supposed to be indivisible by ancient philosophers until
J. J. Thomson who discovered a more fundamental unit in atom, the
electron, by studying cathode rays in 1897, and explained the
phenomena in the plum-pudding model in which the negatively charged
electrons were distributed in a uniform medium of positive charge.
In 1911, this scenario was challenged by the famous experiment of E.
Rutherford, scattering of $\alpha$-particles off gold atoms in foil,
and a positive charge concentrated in a very small nucleus was
speculated in the Rutherford model.
However, it was a matter of big debate about the nuclear
constitution of atoms although Rutherford postulated that ``{\it
Under some conditions, however, it may be possible for an electron
to combine much more closely with the H nucleus, forming a kind of
neutral doublet}''\cite{1}, and that the doublet (latter called {\it
neutron}) would be inside a nucleus.

What does a little nucleus behave like?
It was G. Gamow who suggested to treat atomic nuclei as little
droplets of incompressible nuclear fluid when he worked at the
University of Copenhagen from 1928 to 1931.\cite{2} This liquid drop
model is still popular even an idea of solid nucleus was also
addressed in 1974,\cite{3} motivated by the fact that the giant
resonance might resemble the vibration of an elastic solid.

It is a scientific miracle of the February 1932. Do remember three
events in that month.
(1) L. Landau at the age of 24 published a paper in which an idea of
gigantic nucleus was presented:\cite{4} ``{\it We expect that this
must occur when the density of matter becomes so great that atomic
nuclei come in close contact, forming one gigantic nucleus.}'' The
paper had two notes: ``Received 7 January 1932'' at the beginning
and ending with ``February 1931, Zurich''.
(2) J. Chadwick published a paper with a title of ``{\it Possible
existence of a neutron}'',\cite{5} which was received on Feb. 17,
1932.
(3) A letter of Chadwick was written on Feb. 24, 1932, and sent to
Bohr, with the discussion of neutron and atom.
We can then conclude that Landau speculated gigantic nucleus before
the discovery of neutron.

The motivation for Landau's writing the paper was of duality.
In the years around 1930, a hot topic was on the star equilibrium in
gravity, particularly the maximum mass\cite{6} of white dwarfs where
electron degenerate pressure stands against gravity. What if a
stellar mass is higher than the maximum one?
Landau was thinking about two points: (1) huge gravitational energy
should be released, and (2) the collapse could stop when atomic
nuclei come in close contact.
Landau's effort seeks to relate this scenario to energy source of
stars, and his answer was that forming gigantic nuclei would be the
origin of stellar energy.
Certainly the answer was wrong, but it is the first time that
mankind recognizes a kind of macroscopic matter (an extension of the
microscopic droplet proposed by G. Gamow) at about nuclear
saturation density, which is surely necessary for us to understand
various extreme phenomena through astronomical observations.

It is worth noting that Landau did care about his idea of gigantic
nucleus although he is famous for his fundamental contributions to
condense matter physics, especially the theory of superfluidity.
According to ``{\it Complete list of L D Landau's works}'' provided
by Aksenteva,\cite{7} Landau published totally six {\em Nature}
papers but three were authored only by himself, listed as following.
\begin{romanlist}[(ii)]
\item L. Landau, ``Origin of stellar energy'', {\it Nat.} 141, 333 (1938)
\item L. Landau, ``The theory of phase transitions'', {\it Nat.} 138, 840 (1936)\\
Brief message of ``ZETF 7 (1937) 19, 627; Phys. Z. Sowj. 11 (1937)
26, 545''
\item L. Landau, ``The intermediate state of supraconductors'', {\it Nat.} 141, 688
(1938)\\
Brief message of ``ZETF 13 (1943) 377; J. Phys. USSR 7 (1943) 99''
\end{romanlist}
The first one was actually based upon that\cite{4} published in
1932, and the latter two were summaries of previous works that might
lead to his Nobel prize in physics in 1962.
Why did Landau address again his idea about gigantic nucleus and
stellar energy?
It was said that Landau was submitting the manuscript to {\em
Nature} in order to stand against his political pressure in 1937.
The paper was published finally in 1938, but Landau was still jailed
(he was in prison from 28 April 1938 to 29 April 1939).
One can then see Landau's interests of stars from this real story.

Gigantic nuclei should certainly be neutron-rich if protons ($p$)
and neutrons ($n$) are supposed to be elementary particles, because
electrons ($e$) are {\em inside} a gigantic nucleus.
One can easily come to the conclusion via considering the chemical
equilibrium of the reaction of
\begin{equation}
 e+p\longleftrightarrow n+\nu_e,
\label{ep}
\end{equation}
where the chemical potential of neutrino ($\nu_e$) is negligible. It
is then understandable that the item of gigantic nucleus was not
popular but soon updated by ``{\it neutron star}'' in the scientific
community after the discovery of neutron.

Is there any astronomical consequence of neutron star? Two
astronomers, Baade \& Zwicky,\cite{8} conceived the idea that
forming neutron stars could be the energy source of supernova rather
than of stars, and they investigated the scenario from cosmic rays
(ions) which ``{\em are expelled from them} (supernovae) {\em at
great speeds}'' although the assumed mechanism to produce a neutron
star (``{\em If neutrons are produced on the surface of an ordinary
star they will `rain' down towards the center if we assume that the
light pressure on neutrons is practically zero}'') was not rigorous.
Hot neutron stars were suggested to exist in the cores of supernovae
and to be detected as X-ray sources.\cite{9}
It was also conjectured that the stored energy in rotation of
neutron stars would power supernova remnants.\cite{10}
Certainly the discovery of radio pulsars was a
breakthrough,\cite{11} and pulsars were soon identified as spinning
neutron stars.\cite{12}

Let's conclude this historical note by addressing the fact that L.
Landau and G. Gamow were two in the group known as ``Three
Musketeers'' at the University of Leningrad. Both of them were
thinking about science-breaking questions there and benefited
greatly from each other in their whole lives. Summarily, around
1930, Gamow proposed a liquid picture for microscopic nucleus, but
Landau further suggested macroscopic nuclear matter inside
gravity-bound objects. It is surprising that such stars had really
been discovered!
However, even after 80 years of research, we are still not very
certain about the real nature of gigantic nuclei/neutron stars.

\vspace{0.2cm}
\noindent%
{\bf 2. Two mistakes made by L. Landau at the age of 23}

Constrained by the scientific culture in 1930s, Landau's conjecture
about the gigantic nucleus had to include some concept incorrect.
Let's discuss and analyze two mistakes made in Landau's paper
published in 1932.\cite{4}

{\em Mistake 1}: Quantum theory should be violated in gigantic
nuclei so that a proton and an electron could be ``{\it very close
together}'' (on a scale of femtometer, with a modern language). We
know that an electron interacting with a proton via electromagnetic
force can only make up a relatively {\em weak} system (scale order
of Angstrom) of hydrogen if they obey quantum mechanism which has
been well tested.

The reason that Landau made this mistake was the ignorance of {\it
weak} interaction at that time.
We now know that weak interaction could convert an electron and a
proton to a new particle, neutron, via a reaction of Eq.(\ref{ep}).

In the standard model of particle physics, developed successfully in
the last century, elemental fermions (quarks and leptons) and gauge
bosons are the building blocks, rather than protons (\{$uud$\}) and
neutrons (\{$udd$\}) that were supposed to be point-like particles
in 1930s.
Therefore the reaction of Eq.(\ref{ep}) is essentially converting a
$u$-quark to a $d$-quark by $e+u\longleftrightarrow d+\nu_e$.
In fact, besides $u$ and $d$ valence quarks inside a nucleon, there
are totally six flavors of quarks, which can be divided into two
groups according their masses: light flavors of quarks ($u,d,s$)
with low masses and heavy ones ($c,t,b$) with masses $>$ GeV.
The flavors can be changed by weak interaction although flavor
conservation works during strong interaction.
In this regime, an open question could then arise: Can a gigantic
nucleus have strangeness since it would be easier to excite strange
quarks than heavy quarks?

{\em Mistake 2}: Gravity could be the only interaction to determine
the general property and the global structure of gigantic nuclei.
Landau thought then that only gravitational energy release dominates
and that gigantic nuclei should be gravity-bound, so that he wrote
``{\it Thus we can regard a star as a body which has a neutronic
core the steady growth of which liberates the energy which maintains
the star at its high temperature; the condition at the boundary
between the two phases is as usual the equality of chemical
potential}'' in 1937.

The reason that Landau made the second mistake was the ignorance of
{\it strong} interaction at that time.
The strong force could have at least two important consequences
about gigantic nucleus.
(1) Nuclear energy can be released via nuclear reactions, so that
one can come to a correct conclusion: nuclear power is the origin of
stellar energy.
(2) It is possible that a gigantic nucleus would be self-bound by
strong interaction, rather than gravity-bound, as in the case of a
normal nucleus.

In a word, a lack of knowledge about two kinds of microscopic
interactions (weak and strong) resulted in the mistakes made by
Landau 1930s. 80 years later, it is necessary to modify and develop
the science of gigantic nucleus, with the inclusion of strong and
weak forces which play an important role in nuclear and particle
physics, as well as in astrophysics.
We are trying to do in the next section.

\vspace{0.2cm}
\noindent%
{\bf 3. Gigantic Nuclei revised}

Although the gigantic nucleus concept has been developed to be
modern neutron star models, there do exit essential differences
between little nuclei and gigantic ones.
We think at least two major differences here.
(1) Electrons are included in gigantic nuclei but not in little
nuclei because of relatively large scale (the Compton wavelength is
only $h/(m_ec) = 0.24$\AA).
(2) The average density of gigantic nuclei should be supra-nuclear
density (a few nuclear saturation densities) due to gravitational
force, while gravity would be negligible for little nuclei.
These differences may result in a peculiarity of gigantic nucleus.
Quick questions about a gigantic nucleus include: Are there still
three quarks grouped together as in the case of baryon? Does it have
strangeness? Is it still in a liquid state?
After years of researches, we think that a gigantic nucleus is
actually composed of quark clusters with strangeness and is in a
solid state in order to understand different manifestations of
pulsar-like compact stars.\cite{13} Let's explain in \S3 and \S4,
respectively.

The real state of matter in compact stars is certainly relevant to
the QCD phase diagram in terms of temperature $T$ vs. baryon
chemical potential $\mu_{\rm B}$ (or baryon number density), which
is truly a matter of debate in nuclear and particle physics.
In the regime of low temperature, because of the asymptotic freedom
nature, dense matter may change from a hadronic phase to a
deconfined phase as baryon density increases.
But a very serious problem is: can the density of realistic compact
stars be high/low enough for quarks to become deconfined/confined?

We may approach the state of matter in compact stars along two
directions.

{\em An approach from hadronic state} (a bottom-up scenario).
Let's assume that quarks would not be deconfined in pulsars. However
the confined state may not be simply that of hadrons because of
those two differences as discussed above.
We may expect that a light-flavor symmetry (i.e., that of strange
matter, with approximately equal number densities of $u, d$ and $s$
quarks) would be restored when the baryon density and the
length-scale of gigantic nucleus increase.
We know that nucleon is the lightest baryon, but the typical energy
of electrons would be order of $10^2$ MeV (to be order of or even
higher than the mass difference between $s$-quark and $u$/$d$-quark)
if nucleus keeps not to be broken at supra-nuclear density, while
electrons in strange matter is negligible.
Additionally, gigantic nuclei of light-flavor symmetry might be
ground state of cold matter at a few nuclear densities due to strong
color interaction.
Furthermore, the self-bound gigantic nuclei could be energetically
favored because of huge gravitational energy release, as was noted
by Landau 80 years ago.\cite{4}
Therefore, three flavors of quarks could be grouped together to form
a new hadron-like confined state in gigantic nucleus, and this
multi-quark state may move non-relativistically due to large mass
and localized in lattice at low temperature. We may call this new
state of strong-interaction as ``{\em quark cluster}''.

{\em An approach from free quark state} (a top-down scenario).
Among the speculated states in the diagram, considerable theoretical
efforts have been made to explore the Bardeen-Cooper-Schrieffer-like
color-superconducting phases for cold matter at supra-nuclear
density.\cite{14}
This is worth to do if the coupling between quarks in realistic
compact stars is really relatively weak so that perturbative QCD
(pQCD) works.
It is known that pQCD would work reasonably well only for energy
scale above 1 GeV at least, while the quark chemical potential is
only $\sim 0.4$ GeV for typical pulsars (mass $\sim 1.4M_\odot$ and
radius $\sim 10$ km).
It is then possible that the strong interaction between quarks in
compact stars may result also in the formation of {\em quark
clusters}, with a length scale $l_q$ and an interaction energy
$E_q$.
An estimate from Heisenberg's relation gives if quarks are dressed,
with mass $m_q\simeq 300$ MeV,
\begin{equation}
 l_q \sim {1\over \alpha_s} {\hbar c\over m_qc^2}\simeq {1\over
\alpha_s}~{\rm fm},~~~ E_q \sim \alpha_s^2m_qc^2\simeq
300\alpha_s^2~{\rm MeV}.
\label{Eq}
\end{equation}
This is dangerous for the Fermi state of matter since $E_q$ is
approaching and even greater than the potential $\sim 0.4$ GeV if
the running coupling constant $\alpha_s>1$, and a Dyson-Schwinger
equation approach to non-perturbative QCD shows that the color
coupling should be very strong rather weak, $\alpha_{\rm s}\gtrsim
2$ at a few nuclear densities in compact stars.\cite{13}
Quarks would thus be clustered and localized there.

Therefore a quark cluster phase is conjectured in the QCD phase
diagram.\cite{15} That phase should be in a liquid state at high
temperature, but could be in a solid state when thermal kinematic
energy is lower than the residual strong interaction energy between
quark clusters. Pulsars are in the low temperature limit, and they
would then be solid quark-cluster stars, a modified version of
Landau's gigantic nuclei.
A quark-cluster star would be very similar to a metal ball, but
updating ions/nuclei and electromagnetic interaction in the latter
by quark clusters and strong interaction in the former,
respectively.
It is worth noting that, because of small flavor symmetry broken,
the number density of $s$-quarks is a little bit smaller than that
of $u$/$d$-quarks. The total charge of quarks is then positive,
which may result in the formation of crusts around quark-cluster
stars.

What could be a realistic quark-cluster?
We know that $\Lambda$ particles (with structure $uds$) possess
light-flavor symmetry, and one may think that a kind of quark
clusters would be $\Lambda$-like. However, the interaction between
$\Lambda$'s is attractive, that would render more quarks grouped
together.
Motivated by recent lattice QCD simulations of the H-dibaryons (with
structure $uuddss$), a possible kind of quark clusters, H-clusters,
is proposed.\cite{16}
Besides a general understanding of different manifestations of
compact stars, it is shown that the maximum mass of H-cluster stars
(or simply H stars) could be well above $2M_\odot$ under reasonable
parameters.

\vspace{0.2cm}
\noindent%
{\bf 4. Observational supports for the revision?}

What could be the essential differences between normal neutron stars
(the old version of gigantic nucleus, but developed) and solid
quark-cluster stars (the revised version of gigantic nucleus) by the
observations of pulsar-like stars?
We think there are two.
(1) On the {\em surface}, particles are gravity-bound for neutron
stars while matter is self-confined by strong force for
quark-cluster stars.
(2) The equation of state determines the {\em global} structure:
liquid or rigid? soft or stiff?

The {\em surface} difference has two implications: the mass-radius
($M$-$R$) relation of star and the binding energy of particle.
Self-bound quark-cluster stars have non-zero surface density, and
their radii usually increase as masses increase (specially,
$M\propto R^3$ for low-mass quark-cluster stars with negligible
gravity), while for a gravity-bound neutron star the radius
decreases as the mass increases. The peculiar mass-radius relation
of SAX J1808.4-3658 shows that the star could be a quark
star.\cite{17}
Anyway, an intuitive awareness of the difference could be from the
binding energy.

Pulsar-like stars are populated by radio pulsars, and it seems that
all radio sub-pulses are drifting. These clear drifting sub-pulses
suggest the existence of Ruderman-Sutherland-like gap-sparking and
strong binding of particles on pulsar polar caps, but the calculated
binding energy in neutron star models could not be so high. This
problem could be naturally solved in bare quark-cluster star
scenario due to the strong self-bound nature on surface.\cite{18,19}
In addition, the strong surface binding would result in extremely
energetic exploding because the photon/lepton luminosity of a
quark-cluster surface is not limited by the Eddington limit, and
supernova and $\gamma$-ray bursts could then be
photon/lepton-driven.\cite{20,21,22}
Besides, the observation of non-atomic thermal spectra of dead
pulsars may also hint that there might not exist the atmospheres
speculated in neutron star models, but that this observational
feature could be a manifestation of quark-cluster stars.\cite{23}

Let's turn to the {\em global} difference.
As is addressed in the previous section, quark-cluster stars would
be in a global solid state (like ``cooked eggs''), while only crust
is solid for normal neutron stars (like ``raw eggs'').
Spinning rigid body precesses naturally, either freely or by torque,
and the observation of possible precessions of B1821-11\cite{24} and
others could suggest a global solid structure.
It is said that quark stars cannot reproduce glitches detected soon
after discovery of pulsars, but both normal glitch and slow glitch
could be understood in solid quark cluster-star models.\cite{25,26}
A peculiar action of solid compact stars is star-quake, during which
huge gravitational and elastic energy would be released.
Quake-induced energy may power the flares and bursts of soft
$\gamma$-ray repeaters and anomalous X-ray pulsars.\cite{27}

Additionally, we note also here that the state equation of
quark-cluster stars is stiff because of non-relativistic motion of
quark clusters.
It is well known that the state equation of extremely relativistic
particles is soft, and hence it is a conventional idea that quark
matter is soft.
Nevertheless, theoretical calculations\cite{28,29} show that there
is plenty of parameter space for the maximum mass of quark-cluster
stars to be higher than $2M_\odot$, as was measured in binary system
of PSR J1614-2230.

{\em H-nuggets?}
One consequence of quark-cluster phase could be the existence of
quark nuggets composed of quark clusters. We may call those quark
nuggets as H-nuggets if they are of H-clusters. Can we discover
H-nuggets in cosmic rays?

\begin{figure}[th]
\centerline{\psfig{file=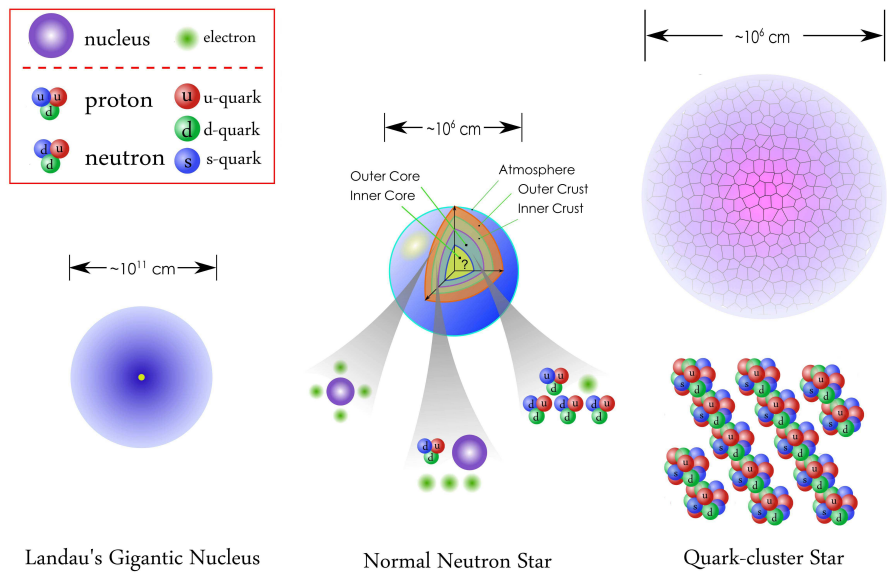,width=12cm}}
 \caption{Three neutron stars, old and new.
 ({\em Thanks to Junwei Yu for his artistic work of drawing.})}
\end{figure}

\noindent%
{\bf 5. Summary}

80 years ago, an idea of gigantic nucleus was presented by Landau,
who tried to understand the origin of stellar energy by combing the
researches of gravity-bound star and Gamow's speculation of nuclear
droplet.
That idea develops then, especially after the discovery of pulsars,
to be very elaborate models of normal neutron stars.
It is also conjectured, from an astrophysical point of
view,\cite{30} that there would exist a quark-cluster state in cold
matter at a few nuclear densities, which could be necessary to
understand different manifestations of pulsar-like compact stars.
Pulsars could then be solid quark-cluster stars.
This concept is shown in Fig.~1.

Let's conclude the paper by a famous sentence of P. W. Anderson
(1923-): ``{\it The ability to reduce everything to simple
fundamental laws does not imply the ability to start from those laws
and reconstruct the universe}''.
As for the state of cold matter at realistic supra-nuclear density,
we get embarrassed and much more difficult because the fundamental
strong interaction law is still uncertain there and would be related
to one of the seven Millennium Prize Problems.

\vspace{0.2cm}
\noindent%
{\bf Acknowledgements.}
%
I would like to thank Dr. Sergey Bastrukov for both historical and
scientific discussions and acknowledge many contributions by members
at the PKU pulsar group. This work is supported by the National
Natural Science Foundation of China (grants 10935001 and 10973002),
the National Basic Research Program of China (grant 2009CB824800),
and the John Templeton Foundation.

\end{document}